# Nonlocal effects on penetration depth of FFLO d-wave superconductors

R. Afzali

Department of Physics, Faculty of Science, KNT University of Technology, P.O. Box 16315-1618 ,Tehran, Iran
E-mail: afzali2001@hotmail.com

PACS numbers: 75.25.Nf

## ABSTRACT

The penetration depth of FFLO d-wave superconductors is calculated, in presence of additional magnetic field for both parallel and perpendicular to the internal FFLO magnetic field, when the nonlocal effects are dominated. The generalized Gorkov equations have been used to obtain the linearized response kernel. It is shown that extra term added to total kernel with no spatially gap is proportional to the momentum of FFLO Cooper pairs, $\vec{Q}$. For the case parallel to the internal FFLO magnetic field, below a crossover temperature $T^*$, it is shown that $(\lambda(T) - \lambda(0))$ is proportional to $T^3$ for both specular reflecting and diffusive boundary. It is noted that both terms in penetration depth with and without $Q$ dependence have the same temperature dependence but magnetic field dependence are different (the terms without and with Q dependence are proportional to $1/h'$ and $1/(h'^2)$ respectively).. Also nonlocal effect on penetration depth of FFLO state in reign $T \gg T^*$ obtains temperature dependence $T^4$. Furthermore, when external magnetic test-field is perpendicular to internal FFLO magnetic field, it is shown that nonlocal effects on penetration depth give the same $Q$ and temperature dependence as the parallel case but with different internal and additional magnetic field dependences.

## I.INTRODUCTION

When the magnetic field is acting on the spin of electrons only, a transition from a normal to a modulated superconducting state or Fulde-Ferrell-Larkin-Ovchinnikov state (FFLO) must occur at low temperatures [1,2]. Recently experiments on CeCoIn$_5$ indicate that in this substance a FFLO state with a modulated order parameter in a strong exchange field is realized[3-5]. Saint-James et al [6] obtained the (T, H) phase diagram for 3D superconductors by assuming that the transition of normal state to FFLO state is of second order. It turns out that the FFLO state only appears at $T < 0.56 T_c$ [6] and that the temperature dependence of the critical magnetic field is strongly influenced by the dimensionality of the system. It is found that the transition from the normal to the FFLO state is of second order in one and two dimensions and is of first order in the three dimensions [7]. The initial instability into the FFLO phase is into a FF phase $(\psi \sim \exp(i\vec{Q}.\vec{X}))$ as opposed to the LO phase $(\psi \sim \cos\vec{Q}.\vec{X})$ that is typically encountered [5]. For unconventional superconductors, impurities induce a change in the structure of the FFLO phase. In the clean limit the FFLO phase is described by an order parameter of from $\cos\vec{Q}.\vec{X}$ while impurities stabilize a $\exp(i\vec{Q}.\vec{X})$ type order parameter [7].

A superconducting state, accompany with an order parameter that oscillates spatially, (FFLO) may be stabilized by a large applied magnetic field or an internal exchange field. Such a FFLO state was subsequently shown to be readily destroyed by impurities and has never been observed in conventional Low-$T_c$ superconductors [9]. The realization of the superconducting state where its order parameter varies spatially is extremely difficult in the classical s-wave type II superconductors. Because the quasiparticle mean free path of superconductor has to be much longer than the coherence length and also the superconductor has to have the Ginzberg-Landau parameter much larger than unity. Indeed these conditions appear to be met in d-wave superconductors like high $T_c$ cuprate and organic superconductors [10]. Reports of existence of FFLO state were done such as organic, heavy-Fermion and high $T_c$ cuprate superconductors as candidates of FFLO state [7,11-14]. Previous experiments have focused on thermodynamic signatures of possible phase transitions from the BCS to FFLO state and then it is suggested Josephson effect to detect the existence of FFLO states [13]. Also STM and NMR can provide a test for the existence of the FFLO state [13]. We suggest another method for indicating FFLO state namely magnetic method. By using nonlocal effects on penetration depth in FFLO state, one can prove the existence of FFLO state by noting difference of temperature dependence of penetration depth caused by nonlocality effects of FFLO state and other superconductors.

In this paper investigates the temperature dependence of the penetration depth of an applied weak magnetic field in the FFLO state of a d-wave superconductor. In conventional s-wave superconductors, the deviation penetration depth from its zero temperature value, $\Delta\lambda(T)$, is proportional to $\exp(-\Delta/T)$. In a pure d-wave superconductor and the other unconventional superconductor with nodes in the gap, $\Delta\lambda(T)$ have quadratic behavior temperature below a certain temperature $T^*$, due to nonlocality and impurities effects on the superconductors [15,16]. Whereas $\Delta\lambda(T)$ have linear behavior temperature at the London limit. The main result of the paper is that, for FFLO state, $\Delta\lambda(T)$ is proportional to $T^3$ if $T \ll T^*$, and $T^4$ if $T \gg T^*$, due to nonlocality effects. Another main result is related to magnetic fields.

I consider here quasi-two dimensional d-wave FFLO superconductors with gap parameter $\Delta\exp(i\vec{Q}\vec{X})$ and also with $\Delta\cos\vec{Q}.\vec{X}$. One and two-dimensional states are more stable than those, due to minimum of their energy [17]. These are under consideration and will be published elsewhere. Semi-infinite superconductor is considered such that, in the basal plane, a-axis and c-axis are parallel to boundary and b-axis perpendicular to superconductor. Also we consider external weak static magnetic field, $\vec{h}$, is parallel to the c-axis. The direction of penetration and external magnetic field characteristic vector, $\vec{q}$, are along the b-axis. Therefore, the applied magnetic field is parallel to internal FFLO magnetic field. Also second I consider external and internal magnetic field perpendicular to each other and then I bring the results of this case.

The paper is organized as follows. In section (II), we approximately obtain Green functions for FFLO state in the presence of external weak magnetic fields and then response kernel of FFLO superconductor in weak magnetic fields is calculated and then nonlocal effects on kernel is considered. In section (III) we obtain nonlocal effects on penetration depth of FFLO in the presence of external weak magnetic fields. Then in section (IV) we have discussion and conclusions.

## II. RESPONSE KERNEL

To obtain penetration depth, we must calculate response kernel. When external magnetic field, defined by $\vec{A}(\vec{X})$, is weak, then one may write the current, $\vec{J}(\vec{X})$, as follows [18,19]

$$\vec{J}(\vec{X}) = (ie/2m)T\sum_{\omega_n}(\nabla_{\vec{X}'} - \nabla_{\vec{X}})|_{\vec{X}'\to\vec{X}}\left(G^F_{1\uparrow\uparrow}(\vec{X}',\vec{X},\omega_n) + G^F_{1\downarrow\downarrow}(\vec{X}',\vec{X},\omega_n)\right) - (Ne^2/m)\vec{A}(\vec{X}) \quad (1)$$

where $N$ is total density of electrons and $G_{1\sigma\sigma}$ is Green function at first approximation due to weak exernal magnetic field. First of all, Gorkov equations [18] must have been solved to obtain Green functions. For this purpose, $\vec{H}'$ is considered to be initial magnetic field vector, which caused spatial order parameter with $\Delta(\vec{X}) = \Delta\,exp(i\vec{Q}\cdot\vec{X})$ or $\Delta(\vec{X}) = \Delta\,\cos(\vec{Q}\cdot\vec{X})$ in quasi two-dimensional d-wave gap parameter. Meanwhile $\vec{Q}$ is total momentum vector of each Cooper pairs in the FFLO state [1,2] and $\Delta$ is the magnitude of gap parameter.

With different direction of external magnetic field with respect to $\vec{H}'$, results are changed. Therefore first I consider the case in which external magnetic field is parallel to $\vec{H}'$ and $\Delta(\vec{X}) = \Delta exp(i\vec{Q}\cdot\vec{X})$. Hamiltonian of system is [18,19] (Throughout this paper we use units in which $k_B = \hbar = c = 1$)

$$\hat{H} = \int d^3 X \hat{\psi}^\dagger_\alpha(\vec{X}) \left\{ \frac{1}{2m} \left(-i\nabla + e\vec{A}(\vec{X})\right)^2 - \mu \right\} \hat{\psi}_\alpha(\vec{X}) + h' \int \alpha \hat{\psi}^\dagger_\alpha(\vec{X}) \hat{\psi}_\alpha(\vec{X}) d^3 X$$
$$+ \int d^3 X \left[ \Delta(\vec{X})^* \hat{\psi}_\uparrow(\vec{X}) \hat{\psi}_\downarrow(\vec{X}) + \hat{\psi}^\dagger_\downarrow(\vec{X}) \hat{\psi}^\dagger_\uparrow(\vec{X}) \Delta(\vec{X}) \right] \qquad (2)$$

The second term in Hamiltonian is Ziemann effect where $h'(\vec{X})\left(\equiv \mu_0 \vec{H}'(\vec{X})\right)$ is supposed is to be constant and has no spatial dependence. $\vec{A}(\vec{X})$ is due to weak constant external magnetic field and I apply as a perturbation to the system. $\hat{\psi}_\alpha(\vec{X})$ and $\hat{\psi}^\dagger_\beta(\vec{X})$ are annihilation and creation Fermionic field operators respectively that satisfy the usual commutation relations (subscripts $\alpha$ and $\beta$ indicate spin $\uparrow$ or $\downarrow$). By using Hisenberg equations of motion, generalized Gorkov equations, dominating for FFLO state in presence of weak external field, are derived in space and time coordinate and because Hamiltonian of system does not explicitly depend on time, using Fourier transformation of Green Functions, generalized Gorkov equations are strenghtly written in frequency coordinate that of course these can not solve exactly. By using the the same procedure done for BCS Green functions due to perturbation $\vec{A}(\vec{X})$, up to first order approximation, Green function can be considered as [19] $G^F_{\uparrow\uparrow}(\vec{X},\vec{X}',\omega_n) = G^F_{0\uparrow\uparrow}(\vec{X},\vec{X}',\omega_n) + G^F_{1\uparrow\uparrow}(\vec{X},\vec{X}',\omega_n)$ where subscripts zero and one stand for the quantities in and out of the presence of external weak magnetic field respectively. It is mentioned that for general case of non-uniform problem, it is always possible to choose the longitudinal part of $\vec{A}(\vec{X})$ in such a way that first order approximation to the $\Delta(\vec{X})$ vanish. Then we get the following linearized equation such as

$$\left(i\omega_n + (1/2m)\vec{\nabla}^2_{\vec{X}} + \mu - h'\right) G^F_{1\uparrow\uparrow}(\vec{X},\vec{X}',\omega_n) + \Delta(\vec{X}) F^\dagger_1(\vec{X},\vec{X}',\omega_n) = \frac{ie}{2m}(\vec{A}\cdot\vec{\nabla}) G^F_{0\uparrow\uparrow}(\vec{X},\vec{X}',\omega_n) \quad (3)$$

where Coulomb gauge , $\vec{\nabla}\cdot\vec{A} = 0$, was used and also the equations for Green functions out of the presence of external weak magnetic field were used. The following equation has translation invariance, by supposing $h'$ is considered to be constant; therefore Green functions depend on $(X - X')$ (the case $h'$ to be function of $\vec{X}$ under consideration and will be published elsewhere). By considering the Fourier transformation of Green Functions, then we obtain the following algebraic Gorkov equations in $\vec{p}$ and $\omega_n = (2n+1)\pi T$ ($n = 0, \pm 1, \pm 2,...$ are Fermionic Matsubara frequencies) spaces such as

$$\left(i\omega_n - \xi_p - h'\right) G^F_{1\uparrow\uparrow}(\vec{p},\omega_n) + \Delta F^{\dagger F}_1(\vec{p} - \vec{Q},\omega_n) = (-e/2m)\vec{A}\cdot\vec{p}\, G^F_{0\uparrow\uparrow}(\vec{p},\omega_n) \quad , \qquad (4)$$

where $\xi_p \equiv p^2/2m - \mu$. In obtaining Eq. (4), also the condition $|\vec{p}| > |\vec{Q}|$ is used. Now we use Tylor expansion of $F^{\dagger F}_1(\vec{p} - \vec{Q},\omega_n)$ about $\vec{Q}$ in Eq. (4). Then we get the Green function at the first order approximation for example [20]

$$G^F_{1\uparrow\uparrow}(\vec{p},\omega_n) = \left(\frac{-e}{2m}\right) \vec{A}\cdot\vec{p} \frac{(-i\omega_n - \xi_p + h')}{(-i\omega_n - \xi_p + h')(i\omega_n - \xi_p - h') + |\Delta|^2} \left( G^F_{0\uparrow\uparrow}(\vec{p},\omega_n) + \frac{\Delta}{(-i\omega_n - \xi_p + h')} F^{\dagger F}_0(\vec{p},\omega_n) \right) \quad (5)$$

To obtain $G^F_{1\uparrow\uparrow}(\vec{p},\omega_n)$, Green functions of FFLO at zero external magnetic field satisfied in equations of FFLO state, are needed [20]. By using $G^F_{0\uparrow\uparrow}(\vec{p},\omega_n) = G_{00\uparrow\uparrow}(\vec{p},\omega_n) + G_{01\uparrow\uparrow}(\vec{p},\omega_n)$ where $G_{00\uparrow\uparrow}$ is Green function with $\vec{Q} = 0$ and $G_{01\uparrow\uparrow}$ is the deviation of Green functions at the first approximation due to existing small quantity $\vec{Q}$, one obtain for example

$$\left[-i\omega_n - \xi_p + h'\right] F^\dagger_{01}(\vec{p},\omega_n) - \Delta^* G_{01\uparrow\uparrow}(\vec{p},\omega_n) = \Delta^* (\vec{Q}\cdot\vec{\nabla}_{\vec{l}}) G_{00\uparrow\uparrow}(\vec{l},\omega_n) \Big|_{\vec{l}=\vec{p}} \qquad (6)$$

Then one can obtain $F_{01}^{\dagger}$, $G_{01\downarrow\downarrow}$ and $G_{01\uparrow\uparrow}$ from Eqs. (40)-(42), and then substituting the answers into Eqs. (34)-(36), we have the following results for $F_0^{\dagger F}(\vec{p},\omega_n)$, $G_{0\uparrow\uparrow}^F(\vec{p},\omega_n)$ and $G_{0\downarrow\downarrow}^F(\vec{p},\omega_n)$

$$G_{0\uparrow\uparrow}^F(\vec{p},\omega_n) = G_{00\uparrow\uparrow}(\vec{p},\omega_n)\left(1 + \Delta^2 g^*(\vec{p},\omega_n)^2(\vec{Q}\cdot\vec{p}/m)G_{00\uparrow\uparrow}(\vec{p},\omega_n)\right) \tag{7}$$

where $g = 1/(i\omega_n - \xi_p + h')$ and we have supposed that $\Delta = \Delta^*$. From Eqs.(6) - (7), it is seen when $\vec{Q}$ tends to zero, then $G_{0\uparrow\uparrow}^F(\vec{p},\omega_n)$ go to $G_{00\uparrow\uparrow}(\vec{p},\omega_n)$ respectively. By substituting Eqs. (7) in Eq. (5), we obtain the following result for $G_{1\uparrow\uparrow}^F(\vec{p},\omega_n)$

$$G_{1\uparrow\uparrow}^F(\vec{p},\omega_n) \equiv -\frac{e}{2m}\left(G_{00\uparrow\uparrow}^F(\vec{p},\omega_n)\vec{A}\cdot\vec{p}G_{0\uparrow\uparrow}(\vec{p},\omega_n) + F_{00}^{\dagger F}(\vec{p},\omega_n)\vec{A}\cdot\vec{p}F_0^{\dagger}(\vec{p},\omega_n)\right)$$
$$= -\frac{e}{2m}G_{00\uparrow\uparrow}(\vec{p},\omega_n)\vec{A}\cdot\vec{p}G_{00\uparrow\uparrow}(\vec{p},\omega_n)$$
$$\left[\left(1 + \Delta^2 g^*(\vec{p},\omega_n)^2\right) + (\vec{Q}\cdot\vec{p}/m)G_{00\uparrow\uparrow}(\vec{p},\omega_n)g^*(\vec{p},\omega_n)\left(\Delta^2 g^*(\vec{p},\omega_n) + \Delta\right)\right] \tag{8}$$

(similarity is the case for $G_{1\downarrow\downarrow}^F(\vec{p},\omega_n)$). Now by substituting Green functions in Eq. (1), one obtain dimensionless response kernel which defined by $\vec{J}(\vec{q}) = -(Ne^2/m)\tilde{K}_{FFLO}(\vec{q},T)\vec{A}(\vec{q})$, as follows

$$\tilde{K}_{FFLO}(q,T,Q,h') = 1 + 2T\int_{-\infty}^{\infty}d\xi\sum_{n=-\infty}^{\infty}\left\langle\hat{p}_{\parallel}^2\left[\left(\frac{\Delta^2 + (-i\omega_n - \xi_{--})(-i\omega_n - \xi_{+-})}{((-i\omega_n - \xi_{--})(i\omega_n - \xi_{-+}) + \Delta^2)((-i\omega_n - \xi_{+-})(i\omega_n - \xi_{++}) + \Delta^2)}\right)\right.\right.$$
$$\left.\times\left(1 + \frac{-h'}{i\omega_n - \xi_{+-}}\right) + \left(\frac{-i\omega_n - \xi_{--}}{(-i\omega_n - \xi_{--})(i\omega_n - \xi_{-+}) + \Delta^2}\right)\left(\frac{-h'}{i\omega_n - \xi_{+-}}\right)\left(\frac{1}{i\omega_n - \xi_{--}}\right)\right]\right\rangle$$

$$+ 2Tv_F Q\int_{-\infty}^{\infty}d\xi\sum_{n=-\infty}^{\infty}\left\langle\hat{p}_{\parallel}^2\hat{p}\cdot\hat{Q}\Delta^2\left(\frac{-i\omega_n - \xi_{--}}{((-i\omega_n - \xi_{--})(i\omega_n - \xi_{-+}) + \Delta^2)^2}\right)\right.$$
$$\left.\times\left(\frac{-i\omega_n - \xi_{+-}}{(-i\omega_n - \xi_{+-})(i\omega_n - \xi_{++}) + \Delta^2}\right)\left(\frac{1}{-i\omega_n - \xi_{--}} + \frac{1}{-i\omega_n - \xi_{+-}}\right)\left(1 + \frac{-h'}{i\omega_n - \xi_{+-}}\right)\right\rangle \tag{9}$$

where $\xi_{+\pm} = \xi_+ \pm h'$, $\xi_{-\pm} = \xi_- \pm h'$, $\hat{p}_{\parallel}$ is the projection of $\hat{p}$ on the boundary (vector $\vec{p}$ is perpendicular to the $c$-axis), $\langle...\rangle$ means averaging over the circular 2D Fermi surface and $v_F$ is Fermi velocity, and $\hat{q}$ is a unit vector perpendicular to the boundary and gives the direction in which the penetration of the external magnetic field takes place. Also $\vec{A}(\vec{q})\cdot\vec{q} = 0$ is used[19] and $\vec{p}_{\pm} \equiv \vec{p} \pm (1/2)\vec{q}$ is appeared to $\xi_{\pm} = (p_{\pm}^2/2m) - \mu$. It is noted that if $\vec{Q}$ and $h'$ go to zero, $\tilde{K}_{FFLO}(q,T,Q,h')$ goes to BCS kernel. For large $\omega_n$ and $\xi$, the integrand on the right hand side of Eq. (9) behaves like $\omega_n^{-m}$ ($m \geq 2$) when $\omega_n \gg \xi$ and like $\xi^{-m}$ when $\xi \gg \omega_n$. Therefore, the integral over $\xi$ and sum over the frequencies $\omega_n$ are divergent. For avoiding this problem, we must first carry out the summation over the frequencies and then the integration over $\xi$. However, it is possible to avoid the need for carrying out the fairly complicated summation over the frequencies in (9). For this purpose, we have to add and subtract some expression from the integrand of (9) [19]. The integral and sum over the frequencies of the difference terms of the integrands is now rapidly convergent, as a result of this we can change the order of integration and summation.

Now I proceed to take nonlocality effect for FFLO superconductors by evaluating $\delta\tilde{k}_{FFLO}(\tilde{q},T,Q,h') \equiv \tilde{k}_{FFLO}(\tilde{q},T,Q,h') - \tilde{k}_{FFLO}(\tilde{q},T=0,Q,h')$. Then one has

$$\delta\tilde{k}_{FFLO}(\tilde{q},T,Q,h') = -2\int_0^\infty d\omega f(\omega)\left\langle \hat{p}_\parallel^2\left[\operatorname{Re}\left(\Delta^2\Big/\sqrt{(\omega-h')^2-\Delta^2}\left(\Delta^2-(\omega-h')^2+\alpha^2\right)\right)+(\omega\to-\omega)\right]\right\rangle$$

$$+\left(v_F Q/2\right)\int_0^\infty d\omega f(\omega)\left\langle \hat{p}_\parallel^2 \hat{p}\cdot\hat{Q}\left[\operatorname{Re}\left(\Delta^2(h'-\omega)\Big/\left((\omega-h')^2-\Delta^2\right)^{3/2}\left(\Delta^2-(\omega-h')^2+\alpha^2\right)\right)+(\omega\to-\omega)\right]\right\rangle \quad (10)$$

where $\alpha \equiv qv_F\hat{q}\cdot\hat{p}/2 = (\pi/2)\alpha_0\tilde{q}\Delta_0\hat{q}\cdot\hat{p}$, $\tilde{q}\equiv q\lambda_0$, $\alpha_0\equiv\xi_0/\lambda_0$, $\lambda_0$ is zero temperature London penetration depth and $\xi_0$ is coherence length and $f(x)$ is Fermi-Dirac distribution function. To obtain Eq. (10), terms given with higher temperature contribution to the $\Delta\lambda(T)\equiv\lambda(T)-\lambda_0$ or an order $(\alpha_0\tilde{q})^2$, which is supposed to be small ($\xi_0\ll\lambda_0$ or $\alpha_0\tilde{q}\ll 1$), were ignored. Also transformation the Matsubara sum into the real frequency integral is used. Because of the presence of the Fermi-Dirac distribution function in Eq. (10), at low temperatures the main contribution to the integration over frequency comes from the interval $\omega \lesssim T$. Therefore, in the average over the Fermi surface the relevant regions are determined by $|\Delta\pm h'|\leq\omega\lesssim T$ and are located around the nodes of the order parameter. At low temperatures, $\Delta$ in Eq. (10) is supposed to be $\Delta\equiv\Delta(\hat{p})=\Delta_0\phi(\hat{p})\approx\Delta_0\phi'(0)\varphi$ where $\varphi$ is the angular deviation of $\hat{p}$ from the given node direction in the basal plane and measured from one of the nodes of the order parameter and $\Delta_0$ is the maximum gap function. For $d_{x^2-y^2}$-wave, $\Delta$ is $2\Delta_0\varphi$. Furthermore, close to the nodes, the dependences of $\hat{p}_\parallel$ and $\alpha$ on $\varphi$ are negligible [15] and this is similarly true for $\hat{p}\cdot\hat{Q}(\equiv\cos\gamma)$. Then finally we have the following result

$$\delta\tilde{k}_{FFLO}(\tilde{q},T,Q,\zeta)\equiv\delta\tilde{k}_{FFLO}(\tilde{q}=0,T)F(z,\zeta)+Qg(z,\zeta) \quad (11)$$

where $\zeta=h'/T=(h'/T^*)/(t)$, $z=\tilde{q}/t$, $t=T/T^*$, $T^*=\alpha_0\Delta_0$, $\delta\tilde{k}_{FFLO}(\tilde{q}=0,T)=-2\ln 2\alpha_0 t$ ( which is equal to $\delta\tilde{k}_{BCS}(\tilde{q},T)=\delta\tilde{k}_{BCS}(\tilde{q}=0,T)F(z)$ for $d_{x^2-y^2}$-wave superconductors [15]). If $h'/T^*$ is considered 55 [21], then one has

$$F(z,\zeta)\simeq\int_0^\infty dxf(x)\left(\sin^{-1}(1/\zeta)-\sqrt{1-\left(8(x+\zeta)^2/\pi^2 z^2\right)}\tan^{-1}\left(y\Big/\sqrt{1-y^2}\sqrt{1-\left(8(x+\zeta)^2/\pi^2 z^2\right)}\right)\right)$$

$$+\left(\zeta\to-\zeta,\ y\equiv 1/\zeta\to y'\equiv 1/(\zeta-1)\right) \quad (12)$$

where $x=\omega/T$. It is noted that the limitations of integral in Eq. (12) must be changed to correct limitations corresponding to whether z is upper or lower than of $4\zeta/\sqrt{2}\pi$. If integrand of Eq. (12) multiply by $\left(8/\pi^2 z^2 T\ln 2\right)(x+\zeta)$, then one can obtain $g(z,\zeta)$. If $h'/T^*$ is to be less than t, then $F(z,\zeta)$ has different form and also one has $g(z,\zeta)=(2\ln 2/\pi\sqrt{2})(v_F\cos\gamma/\Delta_0)(1/z)[F(z,\zeta)-1]$. By using London limit $(z\to 0)$ and Pippard limit $(z\to\infty)$, we can approximate $F(z,\zeta)$. When $h'/T^*$ is 55 [10], then FFLO state exists; but in other cases one doesn't know that FFLO can exists or not, nevertheless this case also is investigated.

## III. PENETRATION DEPTH

For a specular reflecting boundary, the penetration depth, $\lambda$, relative to the London penetration depth, $\lambda_0$, for FFLO state is [22]( by considering smallness of $\delta\tilde{k}_{FFLO}(\tilde{q},T,Q,\zeta)$)

$$\Delta\lambda_{spec}(T)/\lambda_0\approx(2/\pi)\int_0^\infty d\tilde{q}\left(-\delta\tilde{k}_{FFLO}(\tilde{q},T,Q,\zeta)\Big/(\tilde{q}^2+1)^2\right) \quad (13)$$

we finally have the following results

$$T \ll T^*: \frac{\Delta\lambda_{spec}(T)}{\lambda_0} = \frac{11.04b}{3\pi^2\zeta}T^3 + a\frac{v_F\alpha_0 Q\cos\gamma}{\Delta_0\zeta^2}T^3 \quad ; \quad T \gg T^*: \frac{\Delta\lambda_{spec}(T)}{\lambda_0} = \frac{\pi b}{4\zeta^3}T^4 + a\frac{v_F\alpha_0 Q\cos\gamma}{\Delta_0 h'^4}T^4 \quad (14)$$

where $a$ and $b$ is numerical constants. For diffusive boundary, some constants are changed. It is mentioned that, within first order approximation, if we consider $\Delta\cos\vec{Q}.\vec{X}$, then it is sufficient that $Q$ must be replaced to zero in all equations. More investigations are needed to use above formulae, since FFLO state will occur under strong magnetic field. It is mentioned that if $h'/T^*$ is to be less than t, for $T \ll T^*$, $\Delta\lambda_{spec}(T)/\lambda_0$ is approximately $\left[(c_1+c_2 t^2)\right]+\left[Q(b_1 t+b_2)\right]$ where $c_i$ and $b_i$ are constants. $c_1$ is constant and depends on $h'^2$ i.e. $0.4\zeta^2$ and $c_2$ can be considered independence of $h'$ and is $2\ln 2+0.4A$ (where $A$ is nearly 0.5). Also in this case, for $T \gg T^*$, $\Delta\lambda_{spec}(T)/\lambda_0$ is approximately $\left[c_3(h')+c_4 t\right]+b_3 Q$ where $c_i$ and $b_i$ have numerical values. However, from experimental and theoretical viewpoint, we know that when $h'/T^*$ is to be less than t, it is not occurred FFLO state. Therefore this is unstable d-wave superconductor (not d-wave FFLO state). If this case exist, then we have penetration depth of it, due to nonlocal effects.

Also I consider external magnetic field perpendicular to internal strong magnetic field and test-field and consider not only spin singlet pairing but also triplet pairing. [23-25] Then I obtain Green functions ($G_{\uparrow\downarrow}$ and $G_{\downarrow\uparrow}$) and response kernel by long and straight calculation. In spite to different and complicated Green functions of this case, nevertheless $\Delta\lambda_{spec}(T)/\lambda_0$ exactly has the same temperature dependence (we have only $Q$ dependence term)and the different $h'$ (strong internal magnetic field) dependence i.e. when $T \ll T^* (T \gg T^*)$, $T$ dependence of $\Delta\lambda_{spec}(T)/\lambda_0$ is of order 3 and $1/h'$ dependence is of order 3($T$ dependence of $\Delta\lambda_{spec}(T)/\lambda_0$ is of order 4 and $1/h'$ dependence is of order 5) . This is important because the last terms of Eqs. (14) especially indicate FFLO state. Therefore when direction of external magnetic field changed from parallel to perpendicular to internal magnetic field, temperature dependence does not changed but penetration depth become very smaller than former case(external magnetic field parallel to FFLO magnetic field) and also is proportional to $Q$, weak external magnetic field $h$. Also from the case $T \ll T^*$ to $T \gg T^*$, temperature dependence penetration depth is changed from $T^3$ to $T^4$.

IV. DISCUSSION AND CONCLUSIONS

In this paper, we consider quasi two-dimensional d-wave FFLO state with order parameter $\Delta\exp(i\vec{Q}.\vec{X})$ and $\Delta\cos\vec{Q}.\vec{X}$. Then by obtaining Green functions of FFLO state to first order approximation in terms of $Q$, we calculate response kernel of FFLO. For the order parameter $\Delta\cos\vec{Q}.\vec{X}$, our results express that Green functions and response kernel are independent of $Q$, but for the order parameter $\Delta\exp(i\vec{Q}.\vec{X})$, we have extra term, which is proportional to $Q$. Then we proceed to calculate the deviation of response kernel and penetration depth due to nonlocal effects. Recent literature suggests that probably the FFLO state is a superposition of states with different directions of vector $Q$ and even a cascade of transitions between such states is expected [27]. In such a situation it is probably better to neglect the $Q$-dependence of the Green functions altogether, then one can use the results of penetration depth when $Q$ is zero and nonlocal effects on penetration depth obtain for $T \ll T^* (T \gg T^*), c_1(h'^2)+c_2 t^2 \ (c_3(h')+c_4 t)$. The experimental and theoretical work based on BCS theory [15,28-29] on the penetration depth of d-wave unconventional superconductors indicate that penetration depth at low temperatures are proportional to $T^2$. For FFLO, at $T \ll T^* (T \gg T^*)$, by considering $\Delta\exp(i\vec{Q}.\vec{X})$, one can see that both $\Delta\lambda_{spec}(T)/\lambda_0$ and $\Delta\lambda_{diff}(T)/\lambda_0$ are proportional to $T^3 (T^4)$. $h'$ dependence of $\Delta\lambda_{spec}(T)/\lambda_0$ is interesting and the FFLO term, which depends on $Q$, is proportional to $1/h'^2$ ($1/h'^4$), when $T \ll T^*$ ($T \gg T^*$). When external magnetic field perpendicular to internal strong magnetic field, then temperature and magnetic field dependences of $\Delta\lambda_{spec}(T)/\lambda_0$ is the same and it is proportional to $Q$, that especially indicate FFLO state. For this case, when $T \ll T^*$ ($T \gg T^*$), $T$ dependence of $\Delta\lambda_{spec}(T)/\lambda_0$ is of order 3 and $1/h'$ dependence is of order 3 ($T$ dependence of $\Delta\lambda_{spec}(T)/\lambda_0$ is of order 4 and $1/h'$ dependence is of order 5). Also from the case $T \ll T^*$ to $T \gg T^*$, temperature dependence penetration depth is changed from $T^3$ to $T^4$. More investigations are needed to use relations of $\Delta\lambda(T)$. To clarify $\Delta\lambda(T)$, we must have knowledge about $Q$. $Q$ is related to $h'$ and relations between them were given in Ref. [1]. If $\Delta\lambda(T)$ is indicated, then we can get information about strong magnetic field, $h'$ by using Eq. (14). As mentioned above, we obtain the temperature dependence of $\Delta\lambda(T)$ is different from the BCS theoretical work with considering nonlocal effects. Previous experimental work to recognize the FFLO superconductors are mostly based on measuring the specific heat [30] of the sample when the state transit from BCS to FFLO or the Josephson effect in FFLO superconductors [13]. Hence precise measurements on the penetration depth of superconductors will be a good probe for recognizing the occurrence of FFLO state when $\Delta\lambda$ transit from $T^2$ to $T^3$ ($T^4$) dependence if $T \ll T^*$ ($T \gg T^*$).

Acknowlegments:

I would to thank Professor Anthony J. Leggett for his help in providing invaluable information in preparation and research for this article.

REFERENCES:

[1] P. Fulde, and R. A. Ferrell, Phys. Rev. A 135,550 (1964).
[2] A. I. Larkin, and Yu. N. Ovchinnikov, Sov. Phys. JETP 20,762 (1965).
[3] H.A. Radovan, N.A. Fortune, T.P. Murphy, S.T. Hannahs, E.C. Palm, S.W. Tozer and D. Hall, Nature 425, 51 (2003).
[4] T. P. Murphy, Donavan Hall, E. C. Palm, S. W. Tozer, C. Petrovic, Z. Fisk, R. G. Goodrich, P. G. Pagliuso, J. L. Sarrao and J. D. Thompson, Phys. Rev. B65, 100514 (2002).
[5] H. Won, K. Maki, S. Haas, N. Oeschler, F. Weickert and P. Gegenwart, Phys. Rev. B 69, 180504(R) (2004)
[6] D. Saint-James, G. Sarma, and E. J. Thomas, Type II superconductivity (Pergamon Press, New York 1969).
[7] A. I. Buzdin, and H. Kachkachi, Phys. Lett. A 225,341(1997).
[8] D. F. Agterberg, and K. Yang, J. Phys.: Condens. Matter 13,9259(2001).
[9] L. G. Aslamazov, Sov. Phys. JEPT 28,773(1969)
[10] H. A. Radovan, N. A. Fortune, T. P. Murphy, S. T. Hannahs, E. C. Palm, S. W. Tozer and D. Hall, *Nature* 425, 51 (2003).
[11] K. Yang, and S. L. Sondhi, Phys. Rev. B 57,8566(1998).
[12] W. E. Pickett, R. Weht, and A. B. Shick, Phys. Rev. Lett. 83,3713(1999).
[13] K.Yang, and D. F. Agterberg, Phys. Rev. Lett. 84,4970(2000).
[14] K. Yang, Phys. Rev. B 63,140511(2001).
[15] I. Kosztin, and A. J. Leggett, Phys. Rev. Lett. 79,135(1997).
[16] J. P. Hirschfeld, and N. Goldenfeld, Phys. Rev. B 48,4219(1993).
[17] H. Shimahara, J. Phys. Soc. JPN. 67,736(1998).
[18] A. L. Fetter, and J. D. Walecka, Quantum Theory of many-particle systems (McGraw Hill, New York,1971).
[19] A. A. Abrikosov, L. P. Gorkov, and I. E. Dzyaloshinski, Methods of quantum field theory in statistical Physics (Prentice-Hall, Englewood Cliffs, N.J.,1963).
[20] My thesis, 2004.
[21] X. Takada, Prog. Theor. Physics 41, 635 (1969).
[22] M. Tinkham, Introduction to superconductivity (McGraw Hill, New York, 1974).

[23] H. Shimahara, Phys. Rev. B 62, 3524 (2000).
[24] H. Shimahara and M. Kohmoto, *Europhys. Lett.* **57,** 247 (2002).
[25] H. Tanaka, H. Kaneyasu and Y. Hasegawa, J. Phys. Soc. Jpn., 76, 024715 (2007).
[26] C. Mora and R. Combescot, Europhys.Lett. 66,833 (2004).
[27]R. Prozorov, R. W. Giannetta, P. Fournier and R. L. Greene, Phys. Rev. Lett. 85,3700(2000).
[28]S. Djordjevic, E. Farber, G. Deutscher, N. Bontemps, O. Durand, and J. P. Contour, Eur. Phys. J. B 25,407(2002).
[29]K. Gloss, R. Modler, H. Schimanski, C. D. Bredl, C. Geibel, F. Steglish, A. I. Buzdin, N. Sato, and T. Komatsubara, Phys. Rev. Lett. 70,501(1993).